\documentclass[12pt]{article}

\usepackage[dvipdfmx]{graphicx} \usepackage{bmpsize}
\usepackage{amsmath, latexsym, amsfonts, amssymb}
\usepackage{mathrsfs,enumerate}
\usepackage{graphics,epsfig,color}
\usepackage{caption}
\usepackage{subcaption}

\newcommand{\bra}{\left\langle}
\newcommand{\ket}{\right\rangle}

\newtheorem{theorem}{Theorem}[section]
\newtheorem{lemma}[theorem]{Lemma}
\newtheorem{prop}[theorem]{Proposition}

\newtheorem{definition}[theorem]{Definition}

\newenvironment{Proof}{\removelastskip\par\medskip   
\noindent{\em Proof}
\rm}{\penalty-20\null\hfill$\square$\par\medbreak}

\newcommand{\un}{\underline}

\newcommand{\q}{{\ensuremath{\mathbf q}} }
\newcommand{\p}{{\ensuremath{\mathbf p}} }

\newcommand{\be}{{\ensuremath{\mathbf e}} }

\newcommand{\cO}{{\ensuremath{\mathcal O}} }
\newcommand{\cP}{{\ensuremath{\mathcal P}} }

\newcommand{\bbE}{{\ensuremath{\mathbb E}} }

\newcommand{\bbN}{{\ensuremath{\mathbb N}} }

\newcommand{\bbQ}{{\ensuremath{\mathbb Q}} }
\newcommand{\bbR}{{\ensuremath{\mathbb R}} }

\newcommand{\bbZ}{{\ensuremath{\mathbb Z}} }

\title{Macroscopic stability of time evolution of Gibbs measures}
\author{Rapha\"el Lefevere\footnote{Universit\'e de Paris, Laboratoire de Probabilit\'es, Statistiques et Mod\'elisation, UMR 8001, F-75205 Paris, France}
and Shin-ichi Sasa\footnote{Department of Physics, Kyoto University, Kyoto 606-8502, Japan}}

\begin{document}

\maketitle
\begin{abstract}

We introduce two properties, macroscopic mixing and transitive mixing, to represent the macroscopic stability of time evolution of Gibbs measures. We claim that these are fundamental properties of macroscopic systems that exhibit relaxation to an equilibrium state. As an illustration, we show that a simple mechanical system on a lattice possesses these two properties. 
\end{abstract}
\section{Introduction}


Modern statistical mechanics was born with the description of macroscopic bodies through probability distributions over the degrees of freedom of atoms. Suppose that a Gibbs measure describes a macroscopic equilibrium state according to statistical mechanics. When a macroscopic parameter is modified suddenly at $t=0$, the system spontaneously goes to another equilibrium state. This behavior has puzzled theoretical physicists and mathematicians. See \cite{LebowitzPenrose} for a review.  The quest to find a solution to this issue has been one of the motivations for developing ergodic theory.  The {\it mixing} property displayed by some strongly chaotic systems has been considered as a natural candidate to explain this observed stability of the description of macroscopic objects.


While the conventional view is reasonable, the mixing property in dynamical system theory is not directly related to the nature of macroscopic systems. Instead, the observed "macroscopic stability" should be formulated to hold {\it typically} in systems evolving according to Hamiltonian dynamics made of a {\it large} number of particles, and even in non-chaotic systems. The central point of our analysis is that the large deviations function of macroscopic observables may be obtained by using very different probability measures on microscopic degrees of freedom. 
As the large deviations function of observables is identified with thermodynamic potentials, we are naturally led to introduce a notion of equivalence between probability measures.
Thus, in this paper, we formulate the {\it macroscopic mixing property} in terms of the large deviation function of macroscopic observables. 


There is another issue with macroscopic stability when a macroscopic parameter is modified again after some time interval $\Delta t$ from the first parameter change at $t=0$. The system goes to the equilibrium state described by the Gibbs measure compatible with the last value of the parameter. When $\Delta t$ is much larger than the relaxation time of macroscopic observable, the system looks like an equilibrium state. We thus expect that the Boltzmann entropy at the final state never goes below the value of the Boltzmann entropy at the time $\Delta t$. This expectation is not proved immediately. As is well-known, it is easy to show that under a Hamiltonian evolution with an initial Gibbs measure (using Liouville's theorem), the Boltzmann entropy of the system cannot become smaller than its initial value. The proof is based on the Gibbs measure (or measures possessing some properties \cite{Lenard} in general) at the initial time. However, the argument cannot be applied to the measure at time $\Delta t$. In other words, we need to characterize systems to understand a stronger monotonicity property of the Boltzmann entropy. In this paper, we introduce the concept of a {\it transitively mixing} Gibbs measure to represent a property that the measure at time $\Delta t$ can be replaced by a Gibbs measure as far as we observe the macroscopic observables. This concept is the second issue argued in this paper. 


Our main message is that macroscopic mixing and transitive mixing are fundamental properties of macroscopic systems exhibiting relaxation to an equilibrium state. The next problem is to check whether physical systems satisfy these properties. It is pretty challenging to study physical Hamiltonian systems. In this paper, as the first step of the study, we show that a simple discrete-time dynamics of non-interacting particles defined on a lattice with disorder possesses these two properties. We emphasize that the two properties are not trivially satisfied for all systems. That is, we can characterize the macroscopically stable systems by the two properties. 


In Section 2, we describe the set-up we want to consider, recall the main definitions and explain how the Gibbs measures are introduced. In Section 3, we explain the issues at stake in more detail and how to solve them. In Section 4, we introduce a simple mechanical model \cite{Lefevere} of non-interacting particles and random obstacles.  In Sections 5 and 6, we explain how the notions we introduce in Section 3 may be used in this model.

\section{Set-up}

By definition, macroscopic bodies are made of a very large number of particles denoted by $N$.  Let $\Sigma$ be the phase space of the microscopic description of the $N$ particles moving in a $d$ dimensional space.  Typically $\Sigma$ is a subset of $(\bbR^{d}\times \bbR^{d})^N$ that describes the set of possible positions and velocities of the $N$ microscopic particles in $d$ dimensions.  A point of $\Sigma$ is denoted by $x=(\underline \q,\underline \p)= (\q_i,\p_i)_{1\leq i\leq N}$.   A microscopic model for phenomena that are observed at the macroscopic scale is basically given by a Hamiltonian 
$$
H(\un\q,\un\p)=\sum_{i=1}^N\frac{\p_i^2}{2}+U(\un\q)
$$
that describes the dynamics of the microscopic particles. The interaction $U$ generally includes some macroscopic external parameters, like the size of a volume enclosing a gas, the value of a magnetic field and so on.  We generically denote the macroscopic parameters by $\xi$ and indicate the dependence of the Hamiltonian on those parameters by $H^\xi$
The deterministic time evolution in $\Sigma$ is described by  an invertible map $U(\cdot,t):\Sigma\to\Sigma$, $t>0$ where  $U(x,t)$ is the solution of Hamilton's equations at time $t>0$ with initial condition $x=(\un\q,\un\p)$. For simplicity we assume that this solution is well-defined at any time $t>0$ and for any $x\in\Sigma$. 

Some functions $(\phi_N^1,\ldots,\phi_N^n)$ over the phase space $\Sigma$ are identified as the observables describing the phenomenon under consideration in an experiment. In an experimental set-up, the value of the microscopic coordinates are unknown and can therefore only be described by probability distributions. Equilibrium distributions are characterized by the fact that those functions are almost constant on large proportion of the phase space. From this probabilistic point of view,  $(\phi_N^1,\ldots,\phi_N^n)$  will obey a law of large numbers with strong concentration properties.  In other words, the functions corresponding to the observables satisfy a large deviations principle with a rate function having a minimum at the experimentally observed value.  

The probability distribution represents the status of our knowledge about the microscopic coordinates given the knowledge of the values taken by macroscopic variables.  In other words, macroscopic phenomena may be used to infer properties of the microscopic world.  As a well-known example, the temperature of an ideal gas  gives the typical speed of the atoms that make up the gas. Another celebrated example is the Einstein relation between the macroscopic diffusion constant and the variance of the displacement of an atom. 


Let us define ${\mathcal {L}}$ the set of absolutely continuous measures with respect to the Lebesgue measures on $\Sigma$. For a set of observables $\phi_N$ and a vector $m\in\bbR^n$, we define the set
\begin{equation}
{\mathcal {L}}_m=\{\nu\in{\mathcal {L}}:\bra \phi_N\ket_{\nu}=m\}.
	\label{defLm}
\end{equation}
To any measure $\nu$ in ${\mathcal {L}}$, we associate its Shannon-Gibbs
entropy 
\begin{equation}
   S_G[\nu]=-k\int dx \;\nu(x)\log \nu(x),
\end{equation}
where $\nu(x)$ denotes the density with respect to the Lebesgue measure.
A Gibbs measure $\rho^m$ is a maximizer of the Shannon-Gibbs entropy under constraints, namely,
\begin{equation}
         S_G[\rho^m]=\sup_{\nu\in{\mathcal {L}_m}}S_G[\nu].
\label{basic-1}
\end{equation}
 We assume that we have a set of good observables $\phi_N$ in the sense that  
\begin{enumerate}
	\item There is a closed set $\cO\subset\bbR^n$ such that ${\mathcal {L}}_m$ is not empty for any $m\in \cO$.
	\item For any $m\in \cO$, there exists at least one $\rho^m$ such that (\ref{basic-1}) holds.
	\item For any $m\in \cO$, the relation 
\begin{equation}
S_G[\rho^m]:=N s(m)+o(N),
\label{density}	
\end{equation}
where $s$ is a continuous function on $\cO$, holds uniformly in $N$.
\end{enumerate}
We call $s$ the thermodynamic entropy density of the system.  We now make the connection with the Boltzmann entropy.  For a given $m\in \bbR^n$ and $\delta>0$ we define the set of microscopic points on which $\phi_N$ take values in a neighbourhood of $m$ :
\begin{equation}
\Gamma^{\phi_N}_\delta(m)=\{y\in\Sigma :\phi_N(y)\in B_\delta(m)\},
	\label{defgamma}
\end{equation}
where $B_\delta(m)=\{z\in\bbR^n:|z-m|\leq \delta\}$.
The density of the Boltzmann entropy is then defined by
\begin{equation}
	s_B(m)=\lim_{\delta\to 0}\lim_{N\to\infty}\frac 1 N\log |\Gamma^{\phi_N}_\delta(m)|
	\label{defbol}
\end{equation}
where $|\Gamma|$ denotes the volume of a set $\Gamma\in\Sigma$.  In most physically relevant cases, the equality $s_B=s$ holds true.
Informally, one may write a general formula for Gibbs measures :
\begin{equation}
\rho(dx)=\frac{1}{Z(\lambda)}e^{-N\lambda\cdot\phi_N(x)} dx, \quad x\in\Sigma,
\label{Gibbs}
\end{equation}
where $\lambda\cdot\phi_N(x)=\sum_{i=1}^n\lambda^i\phi_N^i(x)$ and $\lambda=(\lambda_1,\ldots,\lambda_n)$ is determined by the set of equations 
$$
\left <\phi_N^i\right>_{\rho}=m^i,\quad i=1,\ldots, n,
$$
and $Z(\lambda)=\int_\Sigma e^{-N\lambda\cdot\phi_N(x)}dx$.

We are interested in the following type of processes. Let  $\tau>0$ and two values of the macroscopic parameters $\xi_1$ and $\xi_2$ entering in the definition of the Hamiltonian. We define  dynamics $\tilde U(\cdot,t):\Sigma\to\Sigma$ by
\begin{equation}
\tilde U(x,t)=\left\{
\begin{array}{ll}
U^1(x,t),& 0\leq t<\tau , \\
U^2(U^1(x,\tau),t-\tau),&  \tau\leq t.
\end{array}
\right.
\label{2times}
\end{equation}

\begin{figure}[thb]
\centering
\includegraphics[width=0.4\textwidth]{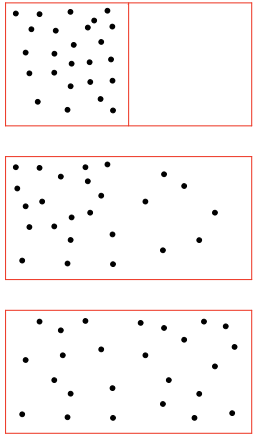}
\caption{An example of  dynamics $\tilde U$. Non-interacting particles are moving freely. On the top figure, they are constrained to move in left part of the volume.  At time $\tau$ the inner central wall is removed and the particles start to occupy the whole available volume.}
\end{figure}
The evolution $U^i$ corresponds to the Hamiltonian $H^{\xi_i}$ and in the sequel we will also use the notation $U_t:=U(\cdot,t)$.
For any $t\geq 0$, and a probability measure $\rho$, we use the following notation:
\begin{equation}
U_t\rho(dx')= \int \delta (U(x,t)-x')\rho(dx).
\label{evolution}
\end{equation}
We obviously have 
$$
\left <f\right>_{\rho(t)}=\left <f(U(\cdot,t))\right >_\rho.
$$

\section{Two issues}
\subsection{Asymptotic equivalence to Gibbs measure}
Let $m\in\bbR^n$ the value taken by the set of observables $\phi_N$ in an initial state and $\rho^m$ a Gibbs measure describing our knowledge of the microscopic degrees of freedom as explained above. Let $\{U(\cdot,t): t>0\}$ a time-evolution described in the previous section. A natural question that arises is whether the measure $\rho^{m}(t)$ becomes close to a (different from $\rho^{m}$) Gibbs measure defined on $\Sigma$ as $t$ becomes larger and larger.  In particular, if $m'=\phi_N(x_{t})$, is $\rho^{m'}$ close in any sense to $\rho^{m}(t)$ as $t\to\infty$? 
The concept of mixing in dynamical systems give an answer to that question. We recall its definition now.  The dynamics $\{U(\cdot,t): t>0\}$ is mixing with respect to the measure $\rho$ if, for any functions $f$ and $g$ in $L^2(\rho)$,
\begin{equation}
	\lim_{t\to\infty}\int_\Sigma f(U(x,t))g(x)\rho(dx)=\int_\Sigma f(x)\rho(dx)\int_\Sigma g(x)\rho(dx).
	\label{usual_mixing}
\end{equation}
If one considers $\mu$ an absolutely continuous probability measure with respect to $\rho$ and replace $g$ by its Radon-Nykodyn derivative then one gets
\begin{equation}
	\lim_{t\to\infty}\int_\Sigma f(U(x,t))\mu(dx)=\int_\Sigma f(x)\rho(dx) .
	\label{usual_mixing2}
\end{equation}

From a physical point of view, however, one needs a notion of equivalence that is weaker.  Indeed, the quantities that can be tested experimentally are derivatives of thermodynamic potentials.  It is therefore natural to base the equivalence between measures on an equality of the thermodynamic potentials computed from $\rho^{m}(t)$ and $\rho^{m'}$.  More precisely, we introduce the {\it macroscopic mixing property}. Let $\mu$ a probability measure over a phase space $\Sigma$ and  $U(\cdot,t): \Sigma\to\Sigma $ the evolution generated by some Hamiltonian dynamics.
We consider the cumulant generating function of the relevant observables $\phi_N$
\begin{equation}
\psi_N(\mu,\eta,U_t)=\frac 1 N\log \left < \exp N\eta\cdot \phi_N(U_t(x))\right>_{\mu},\quad t\geq 0.
\nonumber
\end{equation}
and we use the notation
$$
\psi_N(\mu,\eta):=\psi_N(\mu,\eta,U_0)=
\psi_N(\mu,\eta,{\mathrm {Id}}).
$$
We then introduce the following property:
\begin{definition}
  Let $\rho$ be a Gibbs measure $\rho$ on a phase space $\Sigma$.
  $\rho$  satisfies the {\it macroscopic mixing property } for the dynamics $U(\cdot,t)$ and for a probability measure  $\mu$ if, for any $\eta$,
\begin{equation}
\lim_{t\to\infty}\lim_{N\to\infty}\psi_N(\mu,\eta,U_t)=\lim_{N\to\infty}\psi_N(\rho,\eta).
\label{asymptotic}
\end{equation}
\end{definition}

 The condition (\ref{asymptotic}) is weaker than the usual mixing condition (\ref{usual_mixing}) or (\ref{usual_mixing2}), in particular the large $N$ limit is taken before the $t$ going to infinity.
 We conjecture that this property is true for a large class of systems
 exhibiting relaxation to an equilibrium state.  Indeed in the next section we are able to show that it is satisfied in a simple (and non chaotic)  mechanical system, while it should be noted that it is {\it not} satisfied for Kac ring model \cite{Hiura}.   
 Finally we introduce the useful notations
 \begin{equation}
 m_N(\mu,\eta,U_t)=\nabla_\eta\psi_N(\mu,\eta,U_t), 
 \label{defmN}
 \end{equation}
 and
 $$
 m_N(\mu,\eta,U_0)=m_N(\mu,\eta),\quad m_N(\mu,U_t)=m_N(\mu,0,U_t),
 $$
if the context avoids confusions.

\subsection{Next-day second law}
In the context of the dynamics with a switch of parameters at time $\tau>0$ (see (\ref{2times})), we introduce now a new property, which we call {\it transitive mixing} for Gibbs measures. As we will see in next section, this property is expected to be verified in a large class of dynamical systems that conserve the phase space volume.  But also, it allows to explain that the final value of the entropy is larger than or equal to its value at the time of switching. As detailed in the previous section, a macroscopic system is observed through a set of observables $(\phi_N^1,\ldots,\phi_N^n)$ that are functions defined on the phase space $\Sigma$.  Being ignorant of the exact values of the microscopic coordinates, the observer infers a probability distribution on the microscopic degrees of freedom by choosing the one that maximizes the Shannon entropy under the constraints given by the value of the observables. Those probability distributions are, by definition, the Gibbs measures. The value taken by the Shannon entropy on Gibbs measures is the thermodynamic entropy. The two following facts are essential.

\begin{enumerate}
\item Gibbs measures maximize the Shannon entropy under constraints.
\item The Shannon entropy is constant under a Hamiltonian evolution of the measures.
\end{enumerate}
They are at the root of the proof of the second law of thermodynamics provided by Jaynes \cite{Jaynes,Maes,Goldstein}. Let us first state and prove the second law along those lines.

\begin{prop}
	Let $\{U(\cdot,t):t\geq 0\}$ be a family of phase-space volume preserving maps and $\rho^m$ a Gibbs measure on the phase space $\Sigma$. Let also  
	$$
	m_N(t):=m_N(\rho^m,0,U_t)=\left <\phi_N(U_t(x))\right >_{\rho^m},
	$$
and we assume that for every $t\geq 0$, 
	$$
	\lim_{N\to\infty}m_N(t)=m(t)\in\bbR^n.
	$$
	 Then, if $t>0$ is such that
	\begin{equation}
		\phi_N(x_t)=m(t)+\epsilon_N
		\label{condJ1}
	\end{equation} 
	for some $(\epsilon_N)_N$ such that $\epsilon_N\to 0$ in probability with respect to $\rho^m$, $\forall \eta>0$
	\begin{equation}
		\lim_{N\to\infty}\rho^m[s(m)\leq s(\phi_N(x_t))+\eta]=1. 
	\end{equation} 
\end{prop}
\begin{Proof}
	First, notice that since $U(.,t)$ preserves the phase-space volume, we have 
	\begin{equation}
		S[\rho^m]=S[U_t\rho^m].
	\label{J1}
	\end{equation}
	On the other hand by definition of Gibbs measures, we have that
	\begin{equation}
		S_G[U_t\rho^m]\leq S_G[\rho^{m_N(t)}]
	\label{J2}
	\end{equation}
		since $\bra \phi_N(x)\ket_{U_t\rho^m}=m_N(t)$.
		Thus, using (\ref{density}) we get 
		\begin{eqnarray}
			s(m)&\leq& s(m_N(t))+\frac 1 N o(N).\nonumber\\
		\end{eqnarray}
Since $s$ is continuous, we see that $s(m)\leq s(m(t))$	 and using (\ref{condJ1}) that $s(\phi_N(x_t))$ converges in probability to $s(m(t))$	as $N\to\infty$. This concludes the proof.
\end{Proof}

Let $\{\tilde U(\cdot,t): t\geq 0\}$ the family of volume-preserving transformations of the phase space $\Sigma$ defined in the previous section :
\begin{equation}
\tilde U(x,t)=\left\{
\begin{array}{ll}
U^1(x,t),& 0\leq t<\tau ,\\
U^2(U^1(x,\tau),t-\tau),&  \tau\leq t.
\end{array}
\right.
\label{2times2}
\end{equation}
We assume from now on that we have a well-defined macroscopic evolution in the sense that for every $m\in\cO$ we have that $m_N(\rho^m,\bar U_t)\in\cO$ for every $t>0$ and $N>0$.
We then introduce the following {\it transitive mixing} property.
\begin{definition} Let $m\in\cO$ and $\tilde \rho$ a Gibbs measure, $\tilde\rho$ is transitively mixing for Gibbs measures with respect to $\tilde U$ if for any $m\in\cO$,
  
	$$
	\lim_{t\to\infty}\lim_{N\to\infty}m_N(\rho^m,\tilde U_t)=\lim_{t\to\infty}\lim_{N\to\infty}m_N(\rho^{m_N(\tau)}, U^2_{t-\tau})=\tilde m
	$$
	where $m_N(\tau):=m_N(\rho^m,\tilde U_{\tau})$ and  $\tilde m=\lim_{N\to\infty}m_N(\tilde \rho)$.
\end{definition}
	
This condition is somewhat similar to the autonomy condition (3.3) in \cite{Maes}.
It is however different in several respects. In particular, we do not assume that the dynamics is autonomous with respect to the initial measure but focus instead on the mixing properties of the final equilibrium state.  

\begin{prop}
Let $\tilde\rho$ a measure that is transitively mixing measure for Gibbs measures with respect to $\tilde U$ and let $ m' =\lim_{N\to\infty} m_N(\tau)$. Then,
$$
s(m)\leq s( m')\leq s(\tilde m).
$$	
\end{prop}

\begin{Proof}  We first observe that as in (\ref{J1})
$$
S_G[\rho^{m}]=S_G[U^1_\tau\rho^{m}]\leq S_G[\rho^{m_N(\tau)}].
$$
But we also have that for any $t>\tau$
$$
S_G[\rho^{m_N(\tau)}]= S_G[U^2_{t-\tau}\rho^{m_N(\tau)}_2]\leq S_G[\rho^{m_N(\rho^{m_N(\tau)},U^2_{t-\tau}))}].
$$

Now, we can use the property (\ref{density}) and the continuity of $s$ and take the limit $N\to\infty$, $t\to\infty$  in that order to conclude that 
$$
s(m)\leq s(m')\leq s(\tilde m).
$$

\end{Proof}

\section{ A simple mechanical model}

Consider $\bbZ^2$  and let $\be_1=(1,0)$ and $\be_2=(0,1)$ be the two canonical vectors. We consider particles with velocities in a set
 \begin{eqnarray}
 {\mathcal P}&=&\{-\be_-,\be_-,-\be_+,\be_+\},\quad \be_+=\frac{\be_1+\be_2}{2},\quad \be_-=\frac{\be_1-\be_2}{2},
 \end{eqnarray}
and moving at discrete times on $\bbZ^2$ in a direction given by their velocities. 
While moving, particles may encounter ``mirrors" that change their velocities and thereby deflect their motion. The dynamics is described by a deterministic and invertible map. See (\ref{dynamics1}) for the map which we explain below.
We write the phase space of a single particle as
$$
\hat\Sigma=\bbZ^2\times  {\mathcal P}.
$$  
The absence or presence of a mirror is encoded by a variable $\sigma\in\{0,1\}$,and the action of a mirror on the velocity for any $\be\in {\mathcal P}$ is defined by 
$$
\pi_\sigma(\be)=\sigma \bar \be+(1-\sigma) \be,
$$
where $\overline {\pm \be_+}=\mp{ \be_-}$ and $\overline {\pm \be_-}=\mp{ \be_+}$. When a mirror is present, the velocities are simply reflected then $\pi_1(\be)=\bar\be$. When the mirror is absent, the velocity is not affected  and then $\pi_0(\be)=\be$.
 For $\q\in \bbZ^2$ and $\p\in\cP$, we denote by $\xi(\q+\p)$ the variable that takes the value $1$ if there is a mirror at the site $\q+\p$ and $0$ otherwise.
We then consider a set of variables located on a ``shifted" lattice 
$$
\xi=\{\xi(\q+\be_+)\in\{0,1\}:\q\in\bbZ^2\}.
$$
Then, the particle motion on $\hat\Sigma$ in one time-step is defined by 
\begin{equation}
F_{\xi}(x)=(\q+\p+\pi_{\xi(\q+\p)}(\p),\pi_{\xi(\q+\p)}(\p))
\label{dynamics1}
\end{equation}
for any $x=(\q,\p)\in\Sigma$.  We set 
$$
F_{\xi}(x,1)=F_{\xi}(x)
$$
and for any $t\in\bbN$, $t>1$, we inductively define the iteration of the transformation by
\begin{equation}
	F_{\xi}(x,t)=F_{\xi}\left(F_{\xi}(x,t-1)\right).
\end{equation}

From now on, we will consider the model in a finite box with a horizontal length of $3$, periodic boundary condition in the vertical direction and height of length $R$ in that direction.
Namely, we take 
$$
\Lambda=\{\q=(q_1,q_2)\in\bbZ^2:q_1\in\{1,2,3\}, q_2\in\{1,\ldots, R\}\},
$$ and the phase space describing the motion of $N$ particles is 
$$
\Sigma=\large(\Lambda\times {\mathcal P})^N.
$$
The evolution of the system of the $N$ independent particles is given by :
\begin{equation}
U_{\xi}(\un x,t)=(F_{\xi}(x_1,t),\ldots,F_{\xi}(x_N,t)),\quad \un x=(x_1,\ldots,x_N)\in\Sigma,\quad x_i=(\q_i,\p_i)\in\Lambda\times {\mathcal P}.
\label{dynamics2}
\end{equation}
 The periodic boundary conditions in the vertical direction are enforced by the fact that any value $q_2$ in a pair $\q=(q_1,q_2)$ that appears in the equations below is taken modulo $R$.  The motion of the particles is confined to the box $\Lambda$, namely we impose the following conditions:
\begin{equation}
\xi(\be_++k \be_2)=1,\quad k \in \{0,\ldots , R-1\}
\label{bc1}
\end{equation}
for the left-hand side of the system, and for the right-hand side :
\begin{equation}
\xi(\be_++3\be_1+k \be_2 )=1, \quad k \in \{0,\ldots , R-1\}.
\label{bc2}
\end{equation}
It is easy to check that with those conditions,
$U_{\xi}(\cdot,t)$ is a well-defined map from $\Sigma$ into itself.
We define the rings $R_i,\; i=1,2,3$ by 
$$
R_i=\{(q_1,q_2)\in \Lambda :q_1 =i\}.
$$
The set of relevant observables is the density of particles on each ring
defined by
$$
\phi_N^i(\un\q,\un\p)=\frac 1 R\sum_{j=1}^N{\bf 1}_{R_i}(\q_j),\quad i\in\{1,2,3\}.
$$

We consider a situation in which the $N$ particles are all initially located on the left-most ring $R_1$.  The wall between $R_1$ and $R_2$ is porous, while the wall between $R_2$ and $R_3$ is kept sealed.  At a time $\tau>0$, this
wall is opened and becomes porous. We define the set where the obstacles corresponding to the internal walls are located: 
\begin{equation}
	S=\{\be_++j\be_1+k\be_2:k\in\{0,\ldots,R-1\},j\in\{1,2\}\}.
\end{equation}
To describe this time-dependent process, we choose a random configuration of obstacles $\sigma : S \to \{0,1\}$ distributed as independent Bernoulli random variables with parameter $\gamma$ that represents the density of  mirrors in the system.  We denote by $\bbQ$ the law of $\sigma$.  Next we define a time dependent $\xi(t,\cdot):S\to \{0,1\}$ by 
\begin{equation}
\xi(t,z)=\left\{\begin{array}{ll}
\sigma(z),  & z\in R_1+\be_+ \\
	1,  & z \in R_2 +\be_+,
	
	\end{array}
	\right.
\label{config1}
\end{equation}
if $t<\tau$ and
\begin{equation}
\xi(t,z)=\sigma(z)
	\label{config1}
\end{equation}
for any $z\in S$ if $t\geq \tau$.

Next we define for any $x\in\Lambda\times \cP$
\begin{eqnarray}
	F^1(x,t)&=&F_{\xi(0)}(x,t),\nonumber\\
	F^2(x,t)&=&F_{\xi(\tau)}(x,t),\nonumber\\
	\tilde F(x,t)&=&\left\{
	\begin{array}{ll}
		F^1(x,t),  & 0\leq t \leq \tau\\
		F^2(F^1(x,\tau),t-\tau) ,& \tau < t.
	\end{array}
	\right.
\end{eqnarray}
In a similar way, $\tilde U$ is defined as
\begin{equation}
\tilde U(\underline x,t)=(\tilde F (x_1,t),\ldots,\tilde F (x_N,t))
\label{defut}	
\end{equation}
for $\underline x=(x_1,\ldots,x_N)\in \Sigma$.
\begin{figure}
\centering
\begin{subfigure}{.6\textwidth}
  \centering
  \includegraphics[width=.6\linewidth]{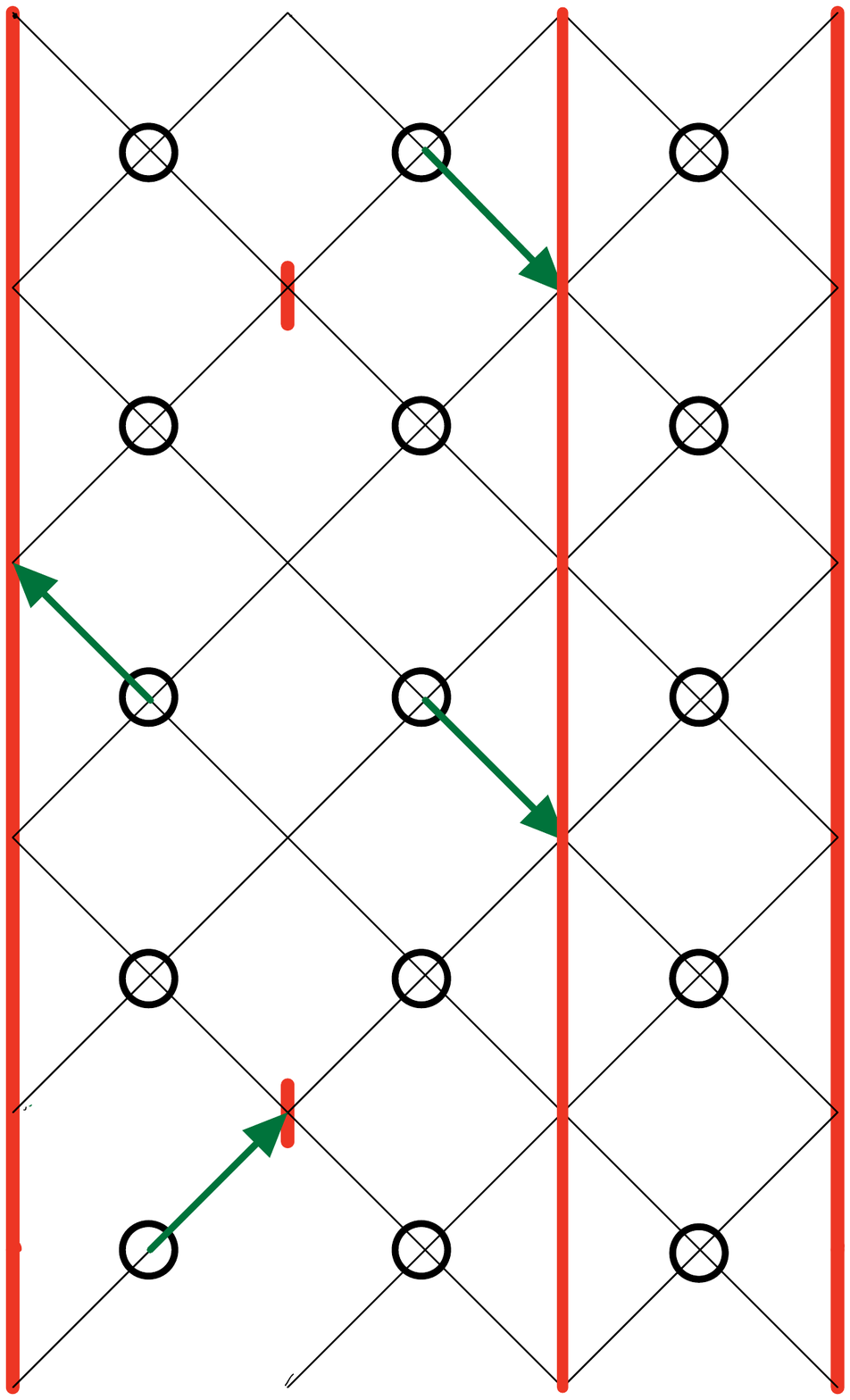}
  \caption{Dynamics $\tilde U$ for $t<\tau$}
  \label{fig:sub1}
\end{subfigure}%
\begin{subfigure}{.6\textwidth}
  \centering
  \includegraphics[width=.6\linewidth]{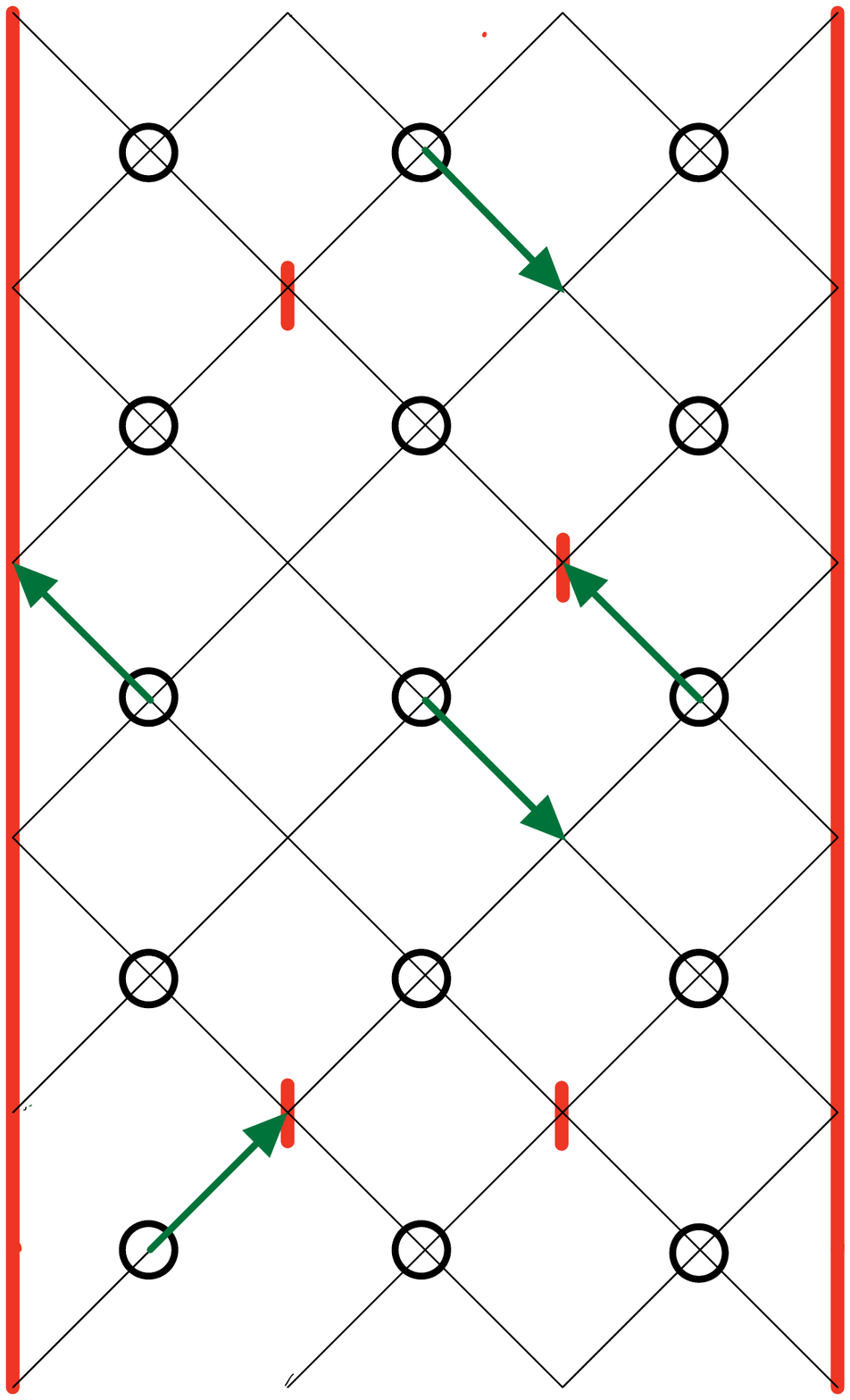}
  \caption{Dynamics $\tilde U$ for $t\geq\tau$}
  \label{fig:sub2}
\end{subfigure}
\label{fig:test}
\end{figure}

For convenience, the number of particles $N$ is always  equal to the number of sites  $R$ on one ring. 
If we start in a situation where all particles are located on ring $R_1$, i.e $\q_i\in R_1$, for $i=1,\ldots,N$ then of course
 $\phi_N^1(\q,\p)=1$ and $\phi_N^i(\q,\p)=0,\;i=2,3$.  We are interested  in the evolution of the expectation of those observables under the evolution (\ref{defut}).
As we consider independent and identically distributed particles, the expectation is always taken with respect to a product measure with identical marginals.  For a given measure $\rho$ on  $\Sigma$ we denote by $\hat\rho$ its single-particle marginal. If no other information is available, the maximum entropy method provides the Gibbs measure for the system of $N$ particles:
$$
\rho_0(\un\q,\un\p)=\prod_{i=1}^N\widehat\rho_0(\q_i,\p_i), \quad \forall (\un\q,\un\p)\in\Sigma,
$$
where
$$
\widehat\rho_0(\q_i,\p_i)=\frac 1 {4R}{\bf 1}_{R_1}(\q_i).
$$
  It is easy to see that 
\begin{equation}
\rho_1(\un\q,\un\p)=\prod_{i=1}^N\frac 1 {8R}{\bf 1}_{R_1\cup R_2}(\q_i)\label{equilibrium1}
\end{equation}
is an invariant probability distribution for the (time-invariant) dynamics  $\tilde U(\cdot,t)$ for $t<\tau$  with {\it any} configuration of obstacles $\xi$.
This probability distribution  is obviously the uniform distribution on the enlarged phase space : $\Sigma_1=((R_1\cup R_2)\times {\mathcal P})^N$.

Let $\Sigma_0=(R_1\times\cP)^N$. Since the dynamics is invertible, it is easy to see that $|U(\Sigma_0,t)|=(4R)^N$ and that $|\Sigma_0|/|\Sigma_1|=(1/2)^N$.  Since $U(\Sigma_0,t)$ is the support of the measure $\rho_0(t)$ and $\Sigma_1$ is the support of $\rho_1$, we see that those two distributions are extremely different.  In spite of that, as far as the the joint fluctuations of the macroscopic observables $\phi_N^1$ and $\phi_N^2$ are concerned, they become more and more similar as time flows (and before the switching time $\tau$). This is the subject of the next section. We finally note that the dynamics $U^1$ is not ergodic with respect to $\rho^1$ : particles moving upward (or downward) do so forever.  Furthermore even if one decomposes the dynamics in two ergodic components, it is not mixing. Indeed each phase space point comes back after a period that is a multiple of $R$ and bounded by the number of phase space points that are accessible. For the dynamics $F^1$, this number is given by $4R$.

\section{ Macroscopic mixing of the dynamics.}
Our goal in this section is to study the asymptotic behavior of the cumulant generating function of the observables $\phi_N=(\phi_N^1,\phi_N^2)$ of the time-evolved measure $U^1_t\rho_0$ and compare it to $\psi_N(\rho_1,\un\eta)$ 
as $N\to\infty$ (and thus $R\to\infty$ since we choose $N= R$) and for $t<\tau$.  Since the dynamical evolution corresponds to $U^1(\cdot,t)$ and depends on $\xi$, $\psi(\rho_0,\eta,U^1_t)$ is therefore itself a random variable through its dependence on $\xi$ (which we don't write explicitly). 
Our goal is to prove that as $N\to \infty$, with  probability  going to 1, the time evolution of the initial Gibbs measure becomes equivalent to the equilibrium Gibbs measure consistent with the macroscopic constraints.  This proves that the system satisfies the property of macroscopic mixing with large probability (of the obstacles distribution).
\begin{theorem}  For any $\gamma\in (0,1)$, there exists $\delta :\bbN\to\bbR$ such that for every $1\leq t< \tau$,
$$
|\delta(t)|\leq \frac 1 2|1-2\gamma|^{\frac{t-1}{2}},
$$ and
for any $\epsilon>0$, any $(\eta_1,\eta_2)\in\bbR^2$,
\begin{equation}
	\lim_{N\to\infty}\bbQ[\left |\psi_N(\rho_0,\eta_1,\eta_2, U^1_t)-\psi_N(\rho_1,\eta_1^+[\delta(t)],\eta_2^-[\delta(t)])\right|>\epsilon]=0,
	\label{convp}
\end{equation}
where 
$$
	\eta^\pm[\delta]=\eta+ \log(1\pm 2\delta).
$$
for any $\delta\in\bbR$.
\label{th1}
\end{theorem}
{\bf Remark}  We will see below that $\psi(\rho_1,\cdot,\cdot)$ is an analytic function of its two arguments.  Therefore by Taylor expanding the $\log$ in $\delta$, it is easy to see that the theorem states that $\psi_N(\rho_0,\eta_1,\eta_2,U^1_t)$ converges in probability to $\psi_N(\rho_1,\eta_1,\eta_2)$. 
\begin{Proof}
We compute first $\psi_N(\rho_1,\eta_1,\eta_2)$.
It is easy to check 
\begin{eqnarray}
\left <e^{N\eta\cdot\phi_N(\un \q)}\right>_{\rho_1}&=&\left <\prod_{i=1}^N e^{(\eta_1{\bf 1}_{R_1}(\q_i)+\eta_2{\bf 1}_{R_2}(\q_i))}\right>_{\rho_1}\nonumber\\
&=& \prod_{i=1}^N\left<e^{(\eta_1{\bf 1}_{R_1}(\q_i)+\eta_2{\bf 1}_{R_2}(\q_i))}\right>_{\rho_1}\nonumber\\
&=&\left(\frac{e^{\eta_1}+e^{\eta_2}}{2}\right)^N.
\end{eqnarray}
Therefore,
\begin{equation}
\psi_N(\rho_1,\eta_1,\eta_2)=\log\left(\frac{e^{\eta_1}+e^{\eta_2}}{2}\right).
\label{psi1}
\end{equation}
We consider now 
\begin{eqnarray*}
	\psi_N(\rho_0,\eta_1,\eta_2,U^1_t)&=&\frac 1 N\log \left <e^{N\eta\cdot \phi_N(\un \q)}\right>_{U^1_t\rho_0}\\
	&=&\log\left<e^{(\eta_1{\bf 1}_{R_1}(\q_1)+\eta_2{\bf 1}_{R_2}(\q_1))}\right>_{U^1_t\rho_0},
\end{eqnarray*}
since $U^1_t\rho_0$ remains product at all time (because the particles do not interact). 
From this expression, we deduce
\begin{equation}
\psi_N(\rho_0,\eta_1,\eta_2,U^1_t)=\log\left(b_1(t)e^{\eta_1}+b_2(t)e^{\eta_2}\right),
\label{obs}
\end{equation}
where 
\begin{equation}
	b_i(t)=\left <{\mathbf 1}_{R_i}(F^1(\cdot,t))\right>_{\widehat\rho_0},\quad i=1,2.
\label{defbi}
\end{equation}
$b_i(t)$ stands for the probability of being on the ring $i$ at time $t>0$ starting with an initial condition uniformly distributed on ring $1$ and for fixed $\xi$.
Note that we have 
\begin{equation}
	b_2(t)=1-b_1(t),
\end{equation}
and that we have dropped the explicit $\xi$ dependence in order to lighten the notations. It is however important to remember that $b_i(t)$ is itself a random variable whose law is determined by $\bbQ$.  We define 
$$
g(x,\eta_1,\eta_2):=\log (x e^{\eta_1}+(1-x)e^{\eta_2}).
$$ 
Using (\ref{obs}), it is easy to check that 
\begin{eqnarray*}
&&g(b_1(t),\eta_1,\eta_2)=\psi(\rho_0,\eta_1,\eta_2,U^1_t), \\
	&& g(\frac 1 2+\delta,\eta_1,\eta_2)=g(\frac 1 2,\eta_1^+[\delta],\eta_2^-[\delta])=\psi(\rho_1,\eta_1,\eta_2),
\end{eqnarray*}
and that $g$ is a Lipschitz continuous function (with constant $C=e^{|\eta_1-\eta_2|}$) as a function of $x$. Therefore, for any function $\delta :\bbN\to\bbR$,
\begin{equation}
|g(b_1(t),\eta_1,\eta_2)-g(\frac 1 2+\delta(t),\eta_1,\eta_2)
| \leq C(\eta_1,\eta_2)|b_1(t)-\frac 1 2-\delta(t)|,
\end{equation}
where $C(\eta_1,\eta_2)=e^{|\eta_1-\eta_2|}$.  We then conclude the proof of the theorem with the help of the following Proposition.
\begin{prop}
For every $\gamma\in (0,1)$, there exists a function $\delta: \bbN\to\bbR$ such that for any $\epsilon>0$ and any $t<\tau$,
\begin{equation}
\lim_{N\to\infty}\bbQ[|b_1(t)-\frac 1 2-\delta(t)|>\epsilon]=0
\label{converge}
\end{equation}
with $|\delta(t)|\leq  \frac 1 2|1-2\gamma|^{\frac{t-1}{2}}$.
\label{propb}
\end{prop}

\end{Proof}
The proof of Proposition \ref{propb} is based on the following Lemma.

\begin{lemma}
There exists a function $\delta: \bbN\to\bbR$ such that, for $t<\min(N,\tau+1)$,
\begin{equation}
\left <b_1(t)\right>_{\bbQ}=\frac 1 2+\delta(t),
\label{convb}
\end{equation}
where $|\delta(t)|\leq  \frac 1 2|1-2\gamma|^{\frac{t-1}{2}}$.
\label{lemma1}
\end{lemma}

\begin{Proof}{\it of Proposition \ref{propb}.}

In view of the Lemma, we see that, using Chebychev's inequality (since $R=N$), it is enough to show that the variance of $b_1(t)$ goes to zero.
We recall that by definition (with the notation $\hat\Sigma_0=R_1\times\cP$) 
\begin{equation}
	b_1(t)=\frac 1{4N}\sum_{x,y\in \hat\Sigma_0}{\mathbf 1}_x(F^1(y,t)).
\end{equation}
Thus taking the variance with respect to $\bbQ$ for $t<\min(\tau,N)$, we obtain 
\begin{eqnarray}
	{\mathrm {Var}}[b_1(t)]&=&\frac 1{16 N^2}\sum_{x,y}{\mathrm {Var}}[{\mathbf 1}_x(F^1(y,t))]\nonumber\\
	&&+\frac 1{16 N^2}\sum_{(x,y)\neq (x',y')}{\mathrm {Cov}}[{\mathbf 1}_x(F^1(y,t)),{\mathbf 1}_{x'}(F^1(y',t))]. 
\end{eqnarray}
The sums run over $x,y,x',y'$ in $\hat\Sigma_0$.
Let us first look at the first sum and let us define the set of phase-space points that can be mapped to the site $x=(\q,\p)$ at time $t$ regardless of the value of $\xi$ 
\begin{equation}
V_t(x)=\{((q'_1,q_2\pm t),\p'): q'_1\in\{1,2\},\p'\in\cP\}.
	\label{setV}
\end{equation}
Whatever the value of $\xi$ and for a given $x$, we have that ${\mathbf 1}_x(F^1(y,t))=0$ if $y\notin V_t(x)$. Thus, for a given $x$, there are at most $4$ terms contributing in the sum over $y\in\hat\Sigma_0$. The variance of a Bernoulli variable being bounded by $1/4$, we get the bound
\begin{eqnarray}
	{\mathrm {Var}}[b_1(t)]\leq\frac 1{16 N}
	+\frac 1{16N^2}\sum_{(x,y)\neq (x',y')}{\mathrm {Cov}}[{\mathbf 1}_x(F^1(y,t)),{\mathbf 1}_{x'}(F^1(y',t))].
\end{eqnarray}

Let us know look at the remaining sum  and examine the dependence of the Bernoulli variables ${\mathbf 1}_x$ on the configuration of $\xi$. From the definition of the dynamics, we see that for a given $x=((q_1,q_2),\p)$ and $t<N$, they depend only on the set of variables $\{\xi(\q''):\q''\in I_t(\q)\}$, 
where
$$
I_t(\q)=\{(\frac 3 2,q_2+\frac 1 2)+j\be_2: j\in \{-t,\ldots,t-1\}\}.
$$
Therefore, whenever $x$ and $x'$ are such that $I_t(\q)\cap I_t(\q')=\emptyset$, we have 
$$
{\mathrm {Cov}}[{\mathbf 1}_x(F^1(y,t)),{\mathbf 1}_{x'}(F^1(y',t))]=0.
$$
Thus, for a given $x$, the sum over $x'$ contains at most $2t$ non-vanishing terms.  On the other hand, the sums over $y$ and $y'$ contribute an overall factor $16$ and each covariance is bounded by the variance of the Bernoulli variable which is at most $1/4$. Therefore, we obtain the bound
\begin{equation}
	{\mathrm {Var}}[b_1(t)]\leq\frac 1{4 N}+\frac{t}{2N}.
	\label{variance0}
\end{equation}
  \end{Proof}

\begin{Proof}{\it of the Lemma.}

The proof basically amounts to show the exponential convergence of a well-chosen Markov chain on which $\left<b_1(t)\right>_{\bbQ}$ depends.
Let us define for $x\in\hat\Sigma_1=(R_1\cup R_2)\times \cP$ and $t\in\bbN$ :
\begin{equation}
	s(t,x):=\left<\left<{\mathbf 1}_{x}(F^1(\cdot,t))\right>_{\widehat\rho_0}\right>_{\bbQ},
\label{defs}
\end{equation}
which has the property that $\sum_{x\in\hat \Sigma_1}s(t,x)=1$ because the dynamics $F^1(\cdot,t)$ is a well-defined map of $\hat\Sigma_1$ onto itself and thus 
$$
\sum_{x\in\hat\Sigma_1}{\mathbf 1}_{x}(F^1(y,t))=1
$$ for any $y\in R_1\times\cP$.
Then, using (\ref{defbi}) and (\ref{defs}) we have 
\begin{eqnarray}
\left <b_i(t)\right>_{\bbQ}&=& \sum_{\p\in\cP}\sum_{\q\in R_i} s(t,\q,\p).
\label{bdecomp}
\end{eqnarray}

We derive now the equations of evolution for $s(t,x)$, for $t<\min(N,\tau+1)$,
\begin{eqnarray}
s(t,x)&=&\sum_{x'\in\hat\Sigma_1}\bbQ[F^1(x',t)= x]\widehat\rho_0(x')\nonumber\\
&=&\sum_{x',x''\in\hat\Sigma_1}\bbQ[F^1(x',t)= x,F^1(x',t-1)= x'']\widehat\rho_0(x')\nonumber\\
&=&\sum_{x',x''\in\hat\Sigma_1}\bbQ[F^1(x'')= x,F^1(x',t-1)= x'']\widehat\rho_0(x')\nonumber\\
&=&\sum_{x',x''\in\hat\Sigma_1}\bbQ[F^1(x'')= x]\bbQ[F^1(x',t-1)= x'']\widehat\rho_0(x')\nonumber\\
&=&\sum_{x''\in\hat\Sigma_1 }K(x;x'')s(t-1,x''),
\label{evol_occup}
\end{eqnarray}
where $K(x;x'')=\bbQ[F^1(x'')= x]$ and $F^1(x,1)=F^1(x)$.
The fourth line follows because by definition $F^1(x'')$ depends on the environment only through the variable $\xi(\q''+\p'')$ and for $t<N$, this variable is independent of $F^1(x',t-1)$. 
Since $F^1(.)$ is a map of $\hat\Sigma_1$ onto itself and 
$$
\bbQ[F^1(x'')= x]=\bbE_{\bbQ}[{\mathbf 1}_x[F^1(x'')]],
$$
it is straightforward to see 
$$
\sum_{x\in\hat\Sigma_1}K(x,x'')=1, \quad \forall x''\in\hat \Sigma_1.
$$
Furthermore, since $F^1(.)$ is invertible on $\hat\Sigma_1$, we  have 
$$
\sum_{x''\in\hat \Sigma_1}K(x,x'')=1, \quad \forall x\in \hat\Sigma_1.
$$
Thus, $K:\hat\Sigma_1\times \hat\Sigma_1\to [0,1]$ is a doubly stochastic kernel.
Next, using the fact that $\bbQ$ is translation-invariant in the ``vertical" direction, we obtain the corresponding property for the transition kernel. Namely, using the notation $e_2=(\be_2,{\bf 0})$, we have 
 
\begin{eqnarray}
K(x+e_2;x'+e_2)&=&K(x;x').\label{trans_inv}
\end{eqnarray}
Using also the fact $K(x,x')=0$ if $|q_2-q_2'|\neq 1$, we see that the transition kernel is entirely specified by 
\begin{equation}
	\bar K(j,\p;i,\p'):=K((j,1+\epsilon(\p'),\p) ;(i,1,\p')), 
	\label{barK}
\end{equation}
where 
$$
\epsilon(\p')=\left\{\begin{array}{cccc}
	1,  & \p'=\be_+ & {\mathrm {or}} & \p'=-\be_- , \\
	-1,  &\p'=\be_- & {\mathrm {or}} &\p'=-\be_+. 
\end{array}
\right.
$$
We give now a representation of the kernel by an $8\times 8$ matrix  $\bar K$. In order to do so, we fix an order on 
\begin{eqnarray*}
M&=&\{1,2\}\times\cP\\
&=&\{(1,\be+),(2,-\be_-),(1,-\be_-),(2,\be_+),(1,\be_-),(2,-\be_+),(1,-\be_+),(2,\be_-)\}\\
&=&\{\sigma_1,\ldots,\sigma_8\}.
\end{eqnarray*}
We then define a matrix $\bar K$ by $\bar K_{ij}=\bar K(\sigma_i,\sigma_j)$
that takes the form

$$
\bar K=\begin{pmatrix}
	A & 0\\
	0 & A
\end{pmatrix},
$$
where 
$$
A=\begin{pmatrix}
	0 & 0 & 1 &0 \\
	0 & 0 & 0 & 1 \\
	\gamma & 1-\gamma &0 & 0\\
	1-\gamma & \gamma & 0 & 0
\end{pmatrix}.
$$
The upper-left block of $\bar K$ corresponds to the upward motion of the particles while lower-right corresponds to the downward motion.

It is straightforward to check that (\ref{trans_inv}) implies that if for some $t\geq 0$, $s(t,\cdot)$ is such that $s(t,x+e_2)=s(t,x)$ for any $x\in\Sigma_1$, then $s(t+1,\cdot)$ has the same property.

Let us take as an initial condition
\begin{equation}
s(0,x)=\left\{
\begin{array}{cc}\frac 1 {4R} & x\in R_1\times\cP\\
0 & x\in R_2\times\cP.
\end{array}
\right.	
\end{equation}
From the above argument, we see that $s(t,x+e_2)=s(t,x)$ for any $t>0$.
We next define a function $\bar s:\bbN\times\{1,2\}\times \cP\to [0,1]$ by 
\begin{equation}
	\bar s(t,i,\p):=R s(t,(i,1),\p).
\label{bbars0}
\end{equation}
The corresponding initial condition for $\bar s$ is 
\begin{equation}
\bar s(0,i,\p)=\left\{
\begin{array}{ll}
\frac 1 4 & i=1,\, \p\in\cP,\\
0 & i= 2,\,\p\in\cP,
\end{array}
\right.
\label{initial}
\end{equation}
and we notice 
\begin{equation}
	\left<b_i(t)\right>_{\bbQ}=\sum_{\p\in\cP}\bar s(t,i,\p)
	\label{bbars}
\end{equation}
with $\left<b_1(0)\right>_{\bbQ}=1$ and $\left<b_2(0)\right>_{\bbQ}=0$. $\bar s(0,\cdot)$ can be represented by a vector $(s(0))_i=s(0,\sigma_i)$ :
\begin{equation}
	\bar s(0):=\frac 1 4\begin{pmatrix}
	u \\ u\\ u\\ u
\end{pmatrix},\quad u=\begin{pmatrix}
	1\\0 
\end{pmatrix}.
\label{barvect}
\end{equation}
From (\ref{evol_occup}), we get the corresponding evolution equation of
$\bar s$, for $t<\min (N,\tau)$,
\begin{equation}
\bar s(t,j,\p)=\sum_{(j'\p')\in M}\bar K(j,\p;j',\p')\bar s(t-1,j',\p')
\label{evol_phi}
\end{equation} 
that can also be written in vector notations
\begin{equation}
\bar s(t)=\bar K \bar s(t-1)=\bar K^{t} \bar s(0).
	\label{evol_phi_mat}
\end{equation}
A little computation shows that
\begin{equation}
	\bar K^2=\begin{pmatrix}
		A^2 & 0\\
		0 & A^2
	\end{pmatrix},\quad A^2=\begin{pmatrix}
B & 0\\
0 & B	
\end{pmatrix},\quad
B=\begin{pmatrix}
		\gamma & 1-\gamma \\
		1-\gamma & \gamma  
	\end{pmatrix}.
\end{equation}
Since $\begin{pmatrix}
		1\\1
	\end{pmatrix}$ and $\begin{pmatrix}
		1\\-1
		\end{pmatrix}$ are eigenvectors of $B$ with eigenvalues $1$ and $1-2\gamma$, it is easy to see 
\begin{equation}
	B^n\begin{pmatrix}
		1\\0
	\end{pmatrix}=\frac{1}{2}\begin{pmatrix}
		1\\1
	\end{pmatrix}+\frac{1}{2}(1-2\gamma)^n\begin{pmatrix}
		1\\-1
		\end{pmatrix}.
\end{equation}
 Thus, for any $n<\frac 1 2\min(R,\tau)$,
  \begin{equation}
 	\bar s(2n,(i,\p))=\frac 1 8 +\epsilon_{2n}(i,\p),
 	\label{even}
 \end{equation}
 where $|\epsilon_{2n}(i,\p)|\leq \frac 1 8|1-2\gamma|^n$.
Thus, we have 
\begin{equation}
  \bar s(2n+1,(i,\p))
  =\sum_{(j,p')\in M}\bar K((i,\p),(j,\p'))\bar s(2n,(j,\p')),
\end{equation}
and since $\bar K$ is obviously a doubly stochastic kernel, we have 
$$
\bar s(2n+1,(i,\p))=\frac 1 8 + \epsilon_{2n+1}(i,\p)
$$
with $|\epsilon_{2n+1}(i,\p)|\leq \frac 1 8|1-2\gamma|^{n}$.
Thus, for any $t<N$,
\begin{equation}
	\bar s(t,(i,\p))=\frac 1 8+\epsilon_t(i,\p)
\end{equation}
with $|\epsilon_t(i,\p)|\leq \frac 1 8|1-2\gamma|^{\frac{t-1}{2}}$.
Using (\ref{bbars}), we see that the claim of the Lemma follows.  
\end{Proof}

We can extend theorem \ref{th1} to the particle densities on each ring.  These are given by the derivatives of the cumulants generating function. Indeed,
\begin{eqnarray}
	m_N(\rho,\eta_1,\eta_2)&=&(\partial_{\eta_1}\psi_N(\rho,\eta_1,\eta_2),\partial_{\eta_2}\psi_N(\rho,\eta_1,\eta_2))\nonumber\\
	&:=&\nabla_\eta\psi(\rho,\eta_1,\eta_2).
\end{eqnarray}

\begin{theorem}  For any $\gamma\in (0,1)$, there exists $\delta :\bbN\to\bbR$ such that, for every $t< \tau$,
$$
|\delta(t)|\leq \frac 1 2|1-2\gamma|^{\frac{t-1}{2}}
$$ and
for any $\epsilon>0$, any $(\eta_1,\eta_2)\in\bbR^2$ and any $t\geq 1$, 
\begin{equation}
	\lim_{N\to\infty}\bbQ[\left\|m_N(\rho_0(t),\eta_1,\eta_2)-m_N(\rho_1,\eta_1^+[\delta(t)],\eta_2^-[\delta(t)])\right\|_\infty>\epsilon]=0.
	\label{convpm}
\end{equation}
where 
$$
	\eta^\pm[\delta]=\eta+ \log(1\pm 2\delta)
$$
for any $\delta\in\bbR$.
\label{th2}
\end{theorem}
\begin{Proof}
	
We have
$$
\nabla_\eta\psi_N(\rho_0(t),\eta_1,\eta_2)=\frac{1}{(b_1(t)e^{\eta_1}+b_2(t)e^{\eta_2})}(b_1(t) e^{\eta_1},b_2(t) e^{\eta_2}).
$$
To finish the proof of the theorem, let us introduce the function $f:\bbR^3\to\bbR^2$, 
$$
f(x,\eta_1,\eta_2)=\left(\frac{x e^{\eta_1}}{xe^{\eta_1}+(1-x)e^{\eta_2}},\frac{(1-x) e^{\eta_2}}{xe^{\eta_1}+(1-x)e^{\eta_2}}\right).
$$
It is straightforward to check that 
\begin{eqnarray*}
	&&\nabla_\eta\psi_N(\rho_1,\eta_1,\eta_2)= f(\frac 1 2,\eta_1,\eta_2), \\
	&&\nabla_\eta\psi_N(\rho_0(t),\eta_1,\eta_2)= f(b_1(t),\eta_1,\eta_2), \\
	&&f(\frac 1 2+\delta,\eta_1,\eta_2)=f(\frac 1 2,\eta_1^+[\delta],\eta_2^-[\delta]),
\end{eqnarray*}
and that each component of $f$ is a Lipschitz continuous function (with constant $C=e^{|\eta_1-\eta_2|}$) as a function of $x$. Therefore, for any function $\delta :\bbN\to\bbR$,
\begin{equation}
\|f(b_1(t),\eta_1,\eta_2)-f(\frac 1 2+\delta(t),\eta_1,\eta_2)
\|_\infty \leq C(\eta_1,\eta_2)|b_1(t)-\frac 1 2-\delta(t)|,
\end{equation}
where $C(\eta_1,\eta_2)=e^{|\eta_1-\eta_2|}$.  We then conclude the proof of the theorem with the help of proposition (\ref{propb}).
\end{Proof}

\section{Next-day second law}

Let us consider now $\rho_2$ the uniform measure on $\Sigma=(R_1\cup R_2\cup R_3)\times\cP$:
$$
\rho_2(\q,\p)=\frac{1}{12 R}, \quad (\q,\p)\in\Sigma.
$$
We want to show that as $N\to\infty$, with large probability (with respect to $\bbQ$), $\rho_2$ is transitively mixing with respect to $\tilde U$.
Let
$$
\cO_2=\{m\in\bbR^3:m^3=0,\quad m^1+m^2=1\}.
$$
Namely, we are going to prove the following theorem.
\begin{theorem}
	For any $m\in \cO_2$, $\tau<t<R$ and for any $\epsilon>0$ :
	\begin{equation}
		\lim_{N\to\infty}\bbQ[|m_N(\rho^m,\tilde U_t)-m_N(\rho_2)|\leq \epsilon+c^{t-\tau}]=1
		\label{next1}
	\end{equation}
	and
	\begin{equation}
		\lim_{N\to\infty}\bbQ[|m_N(\rho^{m_N(\rho^m,U^1_\tau)},U^2_{t-\tau})-m_N(\rho_2)|\leq \epsilon+c^{t-\tau}]=1.
		\label{next2}
	\end{equation}
	
	for some  $0\leq c<1$ and $m_N(\rho_2)=\frac 1 3 (1,1,1)$
\end{theorem}

{\bf Remark} The theorem means that both 
$\lim_{N \to \infty} m_N(\rho^{m_N(\rho^m,U^1_\tau)},U^2_{t-\tau})$
and $\lim_{N \to \infty} m_N(\rho^m,\tilde U_t)$  converge
in probability exponentially fast to  $\lim_{N \to \infty}m_N(\rho_2)$. 

\begin{Proof}
  
	For an arbitrary $m\in \cO_2$, the maximum entropy measure under the condition 
	$$
	\left <\phi_N\right>=(m^1,m^2,0)=m
	$$
	is
	\begin{eqnarray}
		\rho^m(\q,\p)&=&\prod_{j=1}^N\frac{1}{8R}(m^1{\bf 1}_{R_1}(\q_j)+m^2{\bf 1}_{R_2}(\q_j))\nonumber\\
		&:=&\prod_{j=1}^N\hat\rho^m(\q_j,\p_j).
		\label{defrhom}
	\end{eqnarray}
Proceeding as in previous section, we define 

\begin{equation}
	b_i(t)=\left <{\mathbf 1}_{R_i}(\tilde F(\cdot,t))\right>_{\widehat\rho^m},\quad i=1,2,3.
\label{defbi2}
\end{equation}
We see that $m_N(\rho^m(t))=(b_1(t), b_2(t),b_3(t))$ and 
	$$
	\left <b_i(t)\right>_\bbQ=\sum_{\p\in\cP}\bar s(t,i,\p),
	$$
where $\bar s(t,\cdot)$ is defined in a similar ways as in (\ref{bbars0}).
The difference being  that it is  defined on $\{1,2,3\}\times\cP$ and follows an evolution
equation analogous to (\ref{evol_phi})  but with different initial conditions
	$$
	\bar s(0,i,\p)=\left\{\begin{array}
		{ll}
		\frac {m^1}{4} & i=1,\p\in\cP,\\
		\\
		\frac {m^2}{4} & i=2,\p\in\cP,\\
		\\
		0 & i=3, \p\in\cP,
	\end{array}
	\right.
	$$
	or in vector notations using the following ordering of 
	\begin{eqnarray*}
	M&=&\{1,2,3\}\times\cP\\
	&=& \{(1,\be_+),(2,-\be_-),(3,\be_+),(1,-\be_-),(2,\be_+),(3,-\be_-),\\
	&&(1,\be_-),(2,-\be_+),(3,\be_-),(1,-\be_+),(2,\be_-),(3,-\be_+)\}\\
	&=&\{\sigma_1,\ldots,\sigma_{12}\}
\end{eqnarray*}
		
	$$
	\bar s(0)=\frac{1}{4}\begin{pmatrix}
		u \\
		u\\
		u\\
		u
	\end{pmatrix},
	\quad u=\begin{pmatrix}
		m^1\\
		m^2\\
		0
		\end{pmatrix},
	$$
	and
	$$
	\bar s(t)=\left\{\begin{array}{ll}\bar K_1^t \bar s(0),& t\leq \tau,\\
	\\
	\bar K_2^{t-\tau}\bar K_1^\tau \bar s(0), & t > \tau.
	\end{array}
	\right.
	$$
The transition matrix $(\bar K_1)_{ij}=\bbQ[F^1(\sigma_j)=\sigma_i]$ corresponds to the situation where the wall between the rings $R_2$ and $R_3$ is closed while $(\bar K_2)_{ij}=\bbQ[F^2(\sigma_j)=\sigma_i]$ corresponds to the dynamics after the removal of the wall between $R_2$ and $R_3$. We have, 
$$
\bar K_1=\begin{pmatrix}
A_1 & 0 \\
0 & A_1	
\end{pmatrix}
$$
with
$$
A_1=\begin{pmatrix}
	0 & 0 & 0 & 1 & 0 & 0 \\
	0 & 0 & 0 & 0 & 1 & 0 \\
	0 & 0 & 0 & 0 & 0 & 1 \\
	\gamma &1-\gamma &0 &0 &0 &0\\
	1-\gamma &\gamma & 0 & 0 &0 &0\\
	0 & 0 &1 &0 &0 &0
\end{pmatrix}.
$$
We note that 	
$$
A^2_1=\begin{pmatrix}
	B & 0\\
	0 & B
\end{pmatrix},\quad B=\begin{pmatrix}
	\gamma & 1-\gamma & 0\\
	1-\gamma & \gamma & 0\\
	0 & 0 & 1
\end{pmatrix}.	
$$	
By applying iteratively  $\bar K_1$ to $\bar s(0)$, it is easy to check that $\bar s(\tau)=\bar K_1^\tau\bar s(0)$ may be written as
$$
	\bar s(\tau)=\frac{1}{4}\begin{pmatrix}
		u' \\
		u''\\
		u'\\
		u''
	\end{pmatrix},
	\quad{\mathrm {for\; some}}\quad u'=\begin{pmatrix}
		u'_1\\
		u'_2\\
		0
		\end{pmatrix},
		\quad 
		u''=\begin{pmatrix}
		u''_1\\
		u''_2\\
		0
		\end{pmatrix}
	$$
	such that $u'_1+u'_2=u''_1+u''_2=1$. Next, we apply the transition matrix $\bar K_2$ to this new vector. We note that 
	
$$
\bar K_2=\begin{pmatrix}
A_2 & 0 \\
0 & A_2	
\end{pmatrix}
$$
with
$$
A_2=\begin{pmatrix}
	0 & 0 & 0 & 1 & 0 & 0 \\
	0 & 0 & 0 & 0 & \gamma & 1-\gamma \\
	0 & 0 & 0 & 0 & 1-\gamma & \gamma \\
	\gamma &1-\gamma &0 &0 &0 &0\\
	1-\gamma &\gamma & 0 & 0 &0 &0\\
	0 & 0 &1 &0 &0 &0
\end{pmatrix}
$$
such that

$$
A_2^2=\begin{pmatrix}
	C & 0\\
	0 & D
\end{pmatrix},
$$
and
$$ 
C=\begin{pmatrix}
	\gamma & 1-\gamma & 0\\
	(1-\gamma)\gamma & \gamma^2 & 1-\gamma\\
	(1-\gamma)^2 & (1-\gamma)\gamma &\gamma
\end{pmatrix}, \quad
D=\begin{pmatrix}
	\gamma & (1-\gamma)\gamma & (1-\gamma)^2\\
	1-\gamma & \gamma^2 & (1-\gamma)\gamma \\
	0   & 1-\gamma &\gamma
\end{pmatrix}.
$$

$C$ and $D$ are doubly stochastic matrices of a regular Markov chain with a strictly positive and common spectral gap if $\gamma\in (0,1)$.  Applying iteratively $A_2^2$ to $\bar s(\tau)$, we see that we can study separately $C^n u'$ and $D^n u''$.  For any non-zero vector $v=(v_1,v_2,v_2)$ such that $v_1+v_2+v_3=1$, we have $\|C^nv-1/3e\|\leq c^n$ and $\|D^nv-1/3e\|\leq c^n$ for some $c<1$ and $e=(1,1,1)$. 
Thus,
$$
\lim_{n\to\infty}C^n u'=\frac 1 3\begin{pmatrix}
	1\\1\\1
\end{pmatrix},\quad \lim_{n\to\infty}D^n u''=\frac 1 3\begin{pmatrix}
	1\\1\\1
\end{pmatrix}.
$$
Since $A_2$ is a doubly stochastic matrix, we conclude 
$$
\bar s(t)=\frac 1 {12}\begin{pmatrix}
	e\\ e\\ e\\ e\\
\end{pmatrix}+\Delta(t)
$$	
with $|\Delta(t)|<c^{t-\tau}$.  The rest of the proof of (\ref{next1}) amounts to control the variance of $(b_1(t),b_2(t),b_3(t))$. This is similar to the arguments of section 5 and it is left to the reader.

We turn now to the derivation of (\ref{next2}).  Let us define
\begin{equation}
	b_i^m(t-\tau):=m^i_N(\rho^m,U^2_{t-\tau})=\left <{\mathbf 1}_{R_i}(F^2(\cdot,t-\tau))\right >_{\hat\rho^m}.
\end{equation}
Using the notations $m_N(\tau):=m_N(\rho^m,U^1_\tau)$ and $\bar m_N(\tau):=\left <m_N(\rho^m,U^1_\tau)\right>_{\bbQ}$, we have
\begin{eqnarray*}
	&&\left <|b_i^{m_N(\tau)}(t-\tau)-b_i^{\bar m_N(\tau)}(t-\tau)|\right>_\bbQ\\ &\leq& \sum_{\p,\q}\left <{\mathbf 1}_{R_i}(F^2(\p,\q,t-\tau))|\rho^{m_N(\tau)}(\p,\q)-\rho^{\bar m_N(\tau)}(\p,\q)|\right >_\bbQ\\
	&\leq & \sum_{\p,\q}\left (\left <{\mathbf 1}_{R_i}(F^2(\p,\q,t-\tau))\right >_\bbQ\right)^{\frac 1 2}\left (\left <|\rho^{m_N(\tau)}(\p,\q)-\rho^{\bar m_N(\tau)}(\p,\q)|^2\right >_\bbQ\right)^{\frac 1 2}\\
\end{eqnarray*}
by Cauchy-Schwarz inequality.
But using (\ref{defrhom}), we have 
\begin{eqnarray*}
	\left <\left|\rho^{m_N(\tau)}(\p,\q)-\rho^{\bar m_N(\tau)}(\p,\q)\right|^2\right>_{\bbQ}&\leq &\frac 1{4R}\max_{i}{\mathrm {Var}}(m^i_N(\tau)),
\end{eqnarray*}
and thus 
$$
\left |m^i_N(\rho^{m_N(\tau)},U^2_{t-\tau})-m^i_N(\rho^{\bar m_N(\tau)},U^2_{t-\tau})\right |\leq 2 \max_{i}{\mathrm {Var}}(m^i_N(\tau)).
$$
By repeating the argument leading to (\ref{variance0}), it is easy to show that the RHS of the above inequality goes to zero as $N\to\infty$ for $\tau>0$ fixed.  To finish the proof of (\ref{next2}), we write
\begin{eqnarray*}
|m_N(\rho^{m_N(\rho^m,U^1_\tau)},U^2_{t-\tau})-m_N(\rho_2)|&\leq &\left |m_N(\rho^{m_N(\tau)},U^2_{t-\tau})-m_N(\rho^{\bar m_N(\tau)},U^2_{t-\tau})\right|\\&&+\left |m_N(\rho^{\bar m_N(\tau)},U^2_{t-\tau})-m_N(\rho_2)\right|,\\
\end{eqnarray*}
and observe that (\ref{next2}) is proven once we prove that for some $c\in (0,1)$ and for any $\epsilon>0$

	\begin{equation}
		\lim_{N\to\infty}\bbQ[|m_N(\rho^{\bar m_N(\tau)},U^2_{t-\tau})-m_N(\rho_2)|\leq \epsilon+c^{t-\tau}]=1.
	\end{equation}
The proof of this fact is analogous to the proof of (\ref{next1}) and is left to the reader.

\end{Proof}

\section*{Acknowledgements}
R.L. is supported by the ANR-15-CE40-0020-01 grant LSD and benefited from and invitation fellowship from the JSPS.
S. S. is supported by JSPS KAKENHI (Grant Nos. 17H01148, 19H05795, and 20K20425).

\end{document}